\documentclass[prd]{revtex4}%

\usepackage{array}
\usepackage{amsmath}
\usepackage{amssymb}
\usepackage{graphicx}
\usepackage{esint}

\pdfoutput=1

\usepackage{epsfig}
\usepackage{graphics}

\begin{document}
\bigskip
\hskip 4in\vbox{\baselineskip12pt \hbox{FERMILAB-PUB-13-581-A}  }
\bigskip\bigskip\bigskip

\title{ Directional Entanglement of Quantum Fields with Quantum Geometry }

\author{Craig J. Hogan}
\affiliation{  University of Chicago and Fermilab}

\begin{abstract} 
It is conjectured that the spatial structure of  quantum field states is influenced by a new kind of   directional indeterminacy of quantum geometry set by the Planck length, $l_P$, that does not occur in  a classical background geometry.  Entanglement  of fields with geometry modifies the transverse phase of field  states at wavelength $\lambda$ and propagation distance $c\tau$ by about $ \Delta \phi\approx \sqrt{l_P\tau}/\lambda$. The new effect is not detectable in measurements of propagating states that depend only on longitudinal coordinates.     The reduced information content of fields in large systems is consistent with holographic bounds  from gravitation theory,  and  may appear  as measurable quantum-geometrical noise in interferometers.
  \end{abstract}
\maketitle

\section{introduction}

Quantum field theory\cite{wilczek1999} completely accounts for the measured microscopic behavior of matter and its interactions, apart from gravity.  
It blends quantum matter with classical space-time:  field amplitudes are quantized, but    field eigenmodes  have a determinate, continuous spatial structure, embedded in a classical geometry. 
It is widely acknowledged that this  approximation is not a  complete description of nature, because the quantum theory does not include dynamical  degrees of freedom of the geometry. Those are described by a classical theory, general relativity.

In general relativity,  geometry is not a quantum system, but it is a dynamical physical system: it carries  energy and information, and interacts with matter.
 Geometry  couples to an idealized classical model of matter, represented by an  energy-momentum tensor, which is  not a quantum system. 
Although this  approximate description of how matter relates to geometry works well where effects of gravity have been tested, it  also cannot be complete, because real quantum matter cannot be localized in a way that  couples unambiguously to a classical metric.

It is well known that the hybrid of quantum field theory and classical geometry breaks down  in the regime of strong gravity and indeterminate geometry at the Planck scale, $l_P\equiv ct_P\equiv \sqrt{\hbar G/c^3}=1.616\times 10^{-35}$m, where $G$ denotes Newton's constant, $\hbar$ denotes Planck's constant, and $c$ denotes the speed of light--- that is, at very small lengths and high energies.
At  such small scales, the gravity of even a single spatially-localized quantum particle is enough to create a black hole, so   geometry is indeterminate, and the classical approximation to dynamical geometry becomes self-inconsistent. Many theoretical efforts to reconcile geometry with quantum physics have  concentrated on resolving inconsistencies at the Planck scale\cite{Hossenfelder:2012jw}.

However,  the essential  incompatibility of quantum mechanics with dynamical geometry  is not confined to small scales.   Physical quantum states  of radiation  generally  have  indeterminate spatial distributions--- including superpositions of very different space-time histories---  even  on macroscopic scales. It is mathematically impossible to couple a classical space-time to matter in such a state, since there is no scale where the matter has a determinate, classical distribution in space. 

For example, a photon may be radiated from a distant body in a flat space-time in state encompassing a wide angle, and its wave function can expand  to a transverse size of many light years before it is detected; after it is detected, the wave function over its entire path instantly and retroactively  becomes much better localized, to  the  width of the telescope aperture that collected it. 
General relativity can  ignore this kind of indeterminacy in practice because the gravitational effect of such grossly indeterminate states is usually small, but at some level there must be macroscopic quantum properties of geometry that are not included in the standard picture. Somehow, the quantum state of a space-time must be able to consistently couple to indeterminate quantum states of matter, while consistently preserving causal structure, approximate locality, and other classical attributes.

One proposed solution is  that matter and geometry are really a single quantum system\cite{Banks:2009hx,Banks:2009gw,Banks:2010tj,Banks:2011av}.  There are many examples of physical systems where a deeper level of theory reveals completely new degrees of freedom (e.g., \cite{kohn99,laughlin99}).  It could be that matter and geometry seem so different from each other because they emerge with very different kinds of quantum degrees of freedom in  systems much larger than the  Planck scale. The signature properties of classical geometry, such as locality and directionality, could  emerge as a large scale approximate behavior of certain geometrical degrees of freedom. 
Although it is known that general relativity  can  be derived from a statistical or thermodynamic theory \cite{Jacobson:1995ab,Padmanabhan:2009vy,Padmanabhan:2010xh,Verlinde:2010hp}, there is no standard  theory for the relationship between quantum matter and quantum geometrical degrees of freedom.

 A specific new hypothesis concerning  this relationship on scales much larger than the Planck length is proposed here.  Succintly put, the hypothesis is that the directional resolution of any system cannot exceed the limit from Planck diffraction. The number of physically distinct directions in space in a volume of size $L$ is approximately given by $L$ in Planck units, instead of $L^2$ as it would be in a field theory at full Planck resolution.

 It is argued here that the usual, seemingly benign assumption of   spatially-classical field  eigenmodes, whose amplitudes are quantized in field theory, must  break down in a  subtle way on large scales, because it  imposes an unphysical degree of directional independence between quantum field states:  standard field theory implies  higher angular resolution between events than could be achieved by any actual measurement, even with  Planck wavelength radiation.  
 The proposal here is that fields and  geometry behave like  
 subsystems  of a single quantum system, whose directional degrees of freedom are entangled. Field states propagating over a sufficient macroscopic distance are  affected by  Planck scale limits on directional information, and thereby become  entangled with geometrical degrees of freedom originating at the Planck scale.
 Quantum geometry thus affects the behavior of field states in a specific and possibly measurable way, not only at  Planck energy and curvature scales, but also in large systems  in the limit of vanishing gravity. 
  
 The proposed model does not directly address the character of  Planck scale microstates, or systems where gravity is important. The approximations used apply  to  fields propagating on large scales in a nearly-flat space-time,  and to bodies nearly at rest.
Still, the   model implements   nonlocality similar to that  inferred for gravitational systems    (e.g. \cite{Giddings:2004ud,Giddings:2005id,Giddings:2006sj,Hogan:2013aza}),
and matches their maximum information content\cite{Jacobson:1995ab,Padmanabhan:2009vy,Padmanabhan:2010xh,Verlinde:2010hp,Bousso:2002ju}: the  information in a  region  is  given by  the bounding area of a causal diamond  in Planck units, instead of  the volume of phase space as in field theory. 

 Nonlocal and holographic  quantum states of  extended systems are thoroughly studied in some particular highly curved space-times,  for example black holes and anti-de Sitter space (e.g., \cite{Maldacena:1997re,Hawking:2000da,Strominger:2009aj,Nishioka:2009un,Bredberg:2011hp,Maldacena:2013xja,Jensen:2013ora,Sonner:2013mba,Hanada:2013rga}), where precise dualities relate  system states in a curved bulk space to those of a conformal field theory on its boundary.  However, these techniques do  not  address field states in a nearly-flat space-time, which is the subject here, and  is also the regime most accessible to direct experimental tests.

The view  here is that the relationship between  field-like and geometry-like degrees of freedom can be approximately understood from the way a nearly-classical geometry emerges on large scales from a quantum system: the Planck limit  appears, in the emergent space,  in the form of a limit on the  amount of directional information. 
As a result, a field mode in a large system is not a completely separate subsystem from the geometry.
A  theory based on paraxial wave modes is used  to show quantitatively how  Planck-limited directional indeterminacy  entangles quantum fields with geometry  in large systems.
 The  spatial distribution of information in large systems  differs substantially from quantum field theory, in a subtle  but measurable way. 
 
 This proposal also  solves another problem with standard field theory. In a large volume, a standard quantum field system includes states (in general, virtual ones) that have more mass than a black hole of the same size, a contradiction with relativity. 
To solve this problem, Cohen et al.(CKN, ref. \cite{cohen1999})
 proposed an IR bound on the extent of field states, dependent on the UV cutoff of the field system. They proposed  experimental tests based on modifications of field behavior, such as renormalization flow.
The  entanglement  proposed here also results in essentially the same IR bound on field states, but is formulated in a way that depends only on the Planck scale, not on specific properties of the fields. It also predicts specific new observable consequences in  macroscopic systems, since the limit on directional information  degrades the fidelity of angular relationships on large scales in a specific way.

 \section{Spatial Wave Functions of Field States}
 
 A field can be decomposed into different kinds of states that correspond to different kinds of preparation and measurement. 
The most commonly used basis modes are plane waves, which have no uncertainty in orientation but are
completely delocalized in space.  Spatial localization requires a wave packet--- a superposition of  plane wave modes. A wave packet of modes in a single direction can create longitudinal localization, but
  any  transverse localization is associated with transverse momentum, so a localized state necessarily includes some admixture of modes with different orientations.  
 The extreme case is to prepare a state as a nearly point like event, by specifying where a particle is at a  particular place and time. In this case, the subsequent wave function spreads like a spherical wave with an indeterminate direction.

It is possible to define and quantize modes that have some degree of transverse localization, and also some directionality. The spatial character of these states is described not by plane waves, but by paraxial solutions, described below. These states resemble  particles that travel from one somewhat localized place to another, along a somewhat localized path, albeit with some  quantum indeterminacy.  Frequency eigenmodes  have spatial  wave functions that  are confined to a narrow tube out to a certain radial distance, within which they resemble classical paths, and beyond which they spread to a larger angle. 
There is a minimum transverse width at a given distance, the diffraction scale, approximately the geometric mean of wavelength and propagation distance.
A quantum state of a particle with a  direction requires preparation over some finite transverse patch, 
the size of which determines  a propagation distance over which directional information is preserved. At a given distance and frequency, the lowest order paraxial  wave function gives   a minimum intrinsic uncertainty in the direction of a particle's path, which is derived here. In the following section, we will argue that for many purposes, these modes at the Planck scale form a natural macroscopic basis for decomposition of quantum states of emergent geometry.

\subsection{Paraxial Field States and Directional Uncertainty of Paths}

 Techniques of field quantization can be applied to field modes that are not plane waves.
 The appropriate choice of modes depends on boundary conditions that define and prepare the field system.
 Real physical field states are  prepared and measured in a certain way,  with spatially localized interactions.  They can resemble plane waves locally, even over volumes corresponding to many wave lengths, but still have small transverse phase gradients that lead to curved wavefronts at large separations.
One way to discuss this kind of state is to quantize paraxial modes, which split directional from  longitudinal or propagation degrees of freedom in the small angle approximation. Paraxial solutions to the wave equation are familiar from applications in laser cavities\cite{kogelnikli,siegman}.

In three dimensions, consider the amplitude of a component of a  massless field with a sinusoidal time dependence, $A\propto e^{-i\omega t}$, where $\omega =c k= 2\pi c/\lambda$.   
Express the spatial dependence of the field as the  spatial  modulation of a complex carrier wave propagating on the $z$ axis, in the form 
\begin{equation}
A(\vec x)= e^{ik  z} \psi(x,y,z).
\end{equation}
Here $A(\vec x)$ is a complex phasor representing the amplitude and phase at each point, and  Euclidean coordinates $\vec x= x,y,z$ denote position.
The longitudinal coordinate  $z$ corresponds to  position in a particular direction  that defines the orientation of the reference wave, and $x,y$ to positions in  transverse dimensions.
 The field $\psi$ describes deviations of amplitude from a plane wave normal to the $z$ axis. 

Starting with the wave equation for $A$,
\begin{equation}
(\nabla^2 +k^2)A(\vec x)=0,
\end{equation}
the wave equation for $\psi$ becomes
\begin{equation}
\partial^2 \psi/\partial x^2 + \partial^2 \psi/\partial y^2+\partial^2 \psi/\partial z^2 - 2 i k \partial \psi/\partial z = 0.
\end{equation}
  To describe the deviation of the field from a plane wave in the small-angle approximation, assume that the third term is negligible compared with the others:
\begin{equation}\label{paraxial}
{\partial^2 \psi\over \partial x^2 } +{ \partial^2 \psi\over \partial y^2}  - 2 i k {\partial \psi\over\partial z} = 0.
\end{equation}
The paraxial wave equation (Eq. \ref{paraxial}) has the same form as the time-dependent  nonrelativistic Schr\"odinger wave equation in two  dimensions, with $z$ replacing time and $-k$ replacing $m/\hbar$.

Unlike plane waves,   normal modes of this system, called paraxial modes, are spatially confined in the transverse directions. This leads to a  transverse momentum and therefore a transverse spread of propagation direction, described by $\psi$.
The lowest order axially symmetric mode is given by\cite{kogelnikli,siegman}
\begin{equation}
\psi(r,z)=\exp[-i(P+kr^2/2q)],
\end{equation}
where $r^2\equiv x^2+y^2$, $dq/dz=1$, and $dP/dz=-i/q$. The complex beam parameter $q$ can be expressed in terms of two real beam parameters that depend on $z$: the variance $\sigma^2$ of the gaussian profile, and the radius of curvature $R$ of the wave fronts of constant phase, related by:
\begin{equation}
\frac{1}{q} = \frac{1}{R} - \frac{\sqrt{2}i}{k \sigma}.
\end{equation}
In this family of solutions, the gaussian has a minimum width $\sigma_0$ or ``waist''  at $z=0$, where the wave fronts lie in a plane. As a function of $z$, the beam width is
\begin{equation}
\sigma^2(t)= \sigma_0^2[ 1 + \frac{z^2}{k^2\sigma_0^4}],
\end{equation}
and the wavefront curvature radius is
\begin{equation}
R(z)= z[1+ \frac{\sigma_0^4 k^2}{z^2}].
\end{equation}

The width increases, that is, the  beam diverges at larger $z$.  A smaller waist--- that is, a better transverse localization at the origin--- makes for a more rapid divergence, and less localization far away:
\begin{equation}
{\lambda z\over 2 \pi \sigma_0^2}= {2\pi \sigma_0^2\over \lambda R}.
\end{equation}
The wavefronts are nearly flat near the origin ($R=\infty$), and become curved far away, with $R\rightarrow z$ at large $z$. 
The transverse width maintains localization--- suppresses the spreading--- until $z$ is of order $R$.
For $z<<\sigma_0$, a paraxial solution resembles a plane wave, since the width is much larger than the wavelength. For $\sigma_0<z< \sigma_0^2/\lambda$, it resembles a wave confined to a tube of width $\approx \sigma_0$, a ``beam of light''.   For $z>>\sigma_0^2/\lambda$, it resembles a spherical wave emanating from a point source.

Higher order  modes of propagation form a complete and orthogonal set of functions, into which any arbitrary distribution of monochromatic radiation can be decomposed.    Their wave functions  have an  overall Gaussian envelope that follows the fundamental  mode, which represents a lower bound on the overall transverse width. The higher order modes have finer-scale 2D structure in $x$ and $y$ within this envelope. 
 
For any $z$, there is a thus a unique solution that minimizes the width $\sigma(z)$, for which $\sigma/\sigma_0=R/z= \sqrt{2}$. 
The variance of  $\psi(r,z)$  has a minimum value,
\begin{equation}\label{rmin}
\sigma_{min}^2= \langle \Delta r_{min}^2 \rangle =    \lambda z/\sqrt{2}\pi,
\end{equation}
a  transverse ``diffraction  scale'' that grows with system size. 
The corresponding wave function of directional offset $\theta=r/z$ for a  particle path,
 \begin{equation}
\psi(\theta,z)=\exp[-i \theta^2 /2\langle\Delta \theta^2\rangle]
\end{equation}
has a minimum  uncertainty,
\begin{equation}\label{anglemin}
\langle\Delta \theta_{min}^2\rangle = \sigma_{min}^2/z^2= \lambda/ \sqrt{2}\pi z.
\end{equation}
These states display the minimal diffractive  uncertainty inherent in any wave state in a finite system.  Physically,  it may be interpreted as {\it a   directional uncertainty of particle paths, or directional relationships between events, defined by any field at a given wavelength}. States with more precise directionality are possible--- for example, plane waves--- but they have a larger transverse width, so their entire wave functions actually subtend a larger angle, and create  a larger uncertainty for the orientation of  a particle path from the origin to $z$.  It is worth emphasizing once more that this uncertainty is a general property of  frequency eigenstates of any field. Although we are calling it a diffractive uncertainty, it has nothing to do with any additional physical effect of propagation, such as dispersion, scattering or scintillation.

The paraxial modes are  a better approximation than plane waves to  states resembling wave functions of real-world particles that travel from one place to another, since the modes themselves (as opposed to superpositions of modes)  include the maximal amount of  directional localization consistent with a system's size. Their properties thus interpolate between particle and wave. They show how light does not travel in a straight line, or indeed any kind of definite path: it is a wave, and even in flat space travels in a state that is a superposition of straight lines with different orientations.

\subsection{Paraxial Modes as a Basis for Field Quantization}

The paraxial formulation lets us quantize the  amplitude of the field in the usual way as a simple harmonic oscillator, with the usual dependence on classical  space-time coordinates of time and propagation distance. The usual machinery of quantum field theory still applies, such as  raising and lowering operators for occupation number.
This quantization can be applied to the harmonic radial $z$ component, while the directional components  are represented by a  transverse wave function $|\psi\rangle$ that includes explicit geometrical localization and uncertainty, and is time-independent in the paraxial approximation.

Usually,  field modes are quantized by introducing a quantum operator for the amplitude of plane waves. In the plane wave representation, the wave function is
\begin{equation}\label{usual}
\sum_{\vec k, n} \exp[i( \vec k \cdot \vec x- \omega t-\phi)] |A(\vec k, n)\rangle,
\end{equation}
where $|A(\vec k, n)\rangle$ represents a state corresponding to a simple quantum harmonic oscillator with frequency $\omega=c|k|$, and the occupation quantum number $n$ corresponds to particle number in each state. As far as quantum mechanics is concerned, the wave vectors $\vec k$ are also just quantum numbers: they label states of fields in a given volume. The $\vec k$'s depends on the volume of the field system.

For paraxial field modes in the basis aligned with the $z$ axis, we have instead a wave function
\begin{equation}\label{paraxialmodes}
\sum_{k, {\cal K}, n} \exp[i( k  z- \omega t-\phi)] |A(k, n)\rangle |\psi_{\cal K}(x,y;z)\rangle,
\end{equation}
where  $|\psi_{\cal K}(x,y;z)\rangle$ represents a transverse wave function whose eigenstates, labeled by 2D directional quantum numbers ${\cal K}$, have the structure just described, with a diffraction-limited transverse envelope. There is  one radial  wave number $k$, corresponding to the momentum along $z$.  Here, the decomposition depends on the volume as well as a choice of direction, in this case the $z$ axis.
The  number of modes gives the usual answer for total  field degrees of freedom in a volume, but unlike the plane wave decomposition, the individual modes are localized transversely on the diffraction scale.  An excitation of a mode  corresponds to a  particle whose spatial wavefunction is spread spatially over a diffraction-width tube.  The lowest angular modes for a given $z$ correspond to  quantum states of  maximally localized, coherent ``beams of light''.

Provided we include all the higher order modes, there is no physical difference between these descriptions of the overall field system, if the geometry is (as usual)  assumed to be classical. They  simply quantize different decompositions of a wave field, and refer to different ways of preparing and measuring field states.  
 In many situations, the waist size is so much bigger than a wavelength that  for many  purposes in local interactions, these states behave physically in almost the same way as plane waves. However, the paraxial mode decomposition  explicitly displays a physical limit on directional resolution.

\section{Entanglement of Field States with  Quantum Geometry}

\subsection{Planck Limited Directional Information}

Consider a plane wave in a space-time volume  defined by  duration $\tau= z/c$. (That is, the spacetime region invariantly defined by the future and past light cones of the two  events on some world-line, with this proper timelike separation). The  normal to the wavefronts defines a direction with an uncertainty  $\approx \lambda/c\tau$, much less than the diffractive directional uncertainty $\approx \sqrt{\lambda/c\tau}$. Of course, that precision comes at the expense of complete transverse delocalization: there is no transverse position information in the plane wave.
Such a state cannot be used as a basis for   geometrical states that resemble  a  quasi-classical spacetime; these must not have the character of plane waves, because they must appear on large scales as a nearly-classical geometry, with approximate localization in all three spatial dimensions.  Transverse localization is accompanied by  directional uncertainty.  

For geometry, a minimum uncertainty is given by the Planck scale.
The  diffraction limit (Eq. \ref{anglemin}) for a Planck  wavelength state and a propagation distance $z$, 
\begin{equation}\label{planckdiffraction}
\Delta \theta_P= \langle\Delta \theta_{min}^2(\lambda=l_P)\rangle^{1/2} =  (\sqrt{2} l_P/\pi z  )^{1/2},
\end{equation}
   corresponds to the best precision with which directions between events can in principle be measured at separation $z$, using Planck wavelength radiation or particles.  We   propose the hypothesis that  {\it  Eq. (\ref{planckdiffraction})
approximately  represents a fundamental  limit to directional resolution}--- a new effect of quantum geometry.  
 Information in  geometrical degrees of freedom thus has a different character from that in  quantum fields, and coherently affects the spatial structure of field states.

\begin{figure}[t]
 \epsfysize=4in 
\epsfbox{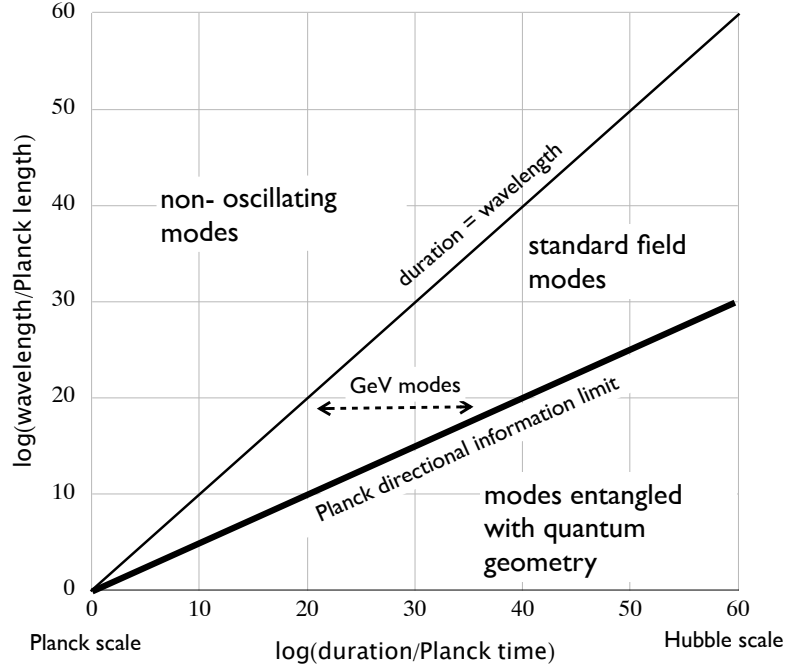} 
\caption{ \label{validity}
Mode wavelengths and system durations for  field modes, showing the scales affected in a geometry that has the directional fidelity of Planck scale waves. Wavelength $\lambda$ is plotted as function of causal diamond duration $\tau$, both expressed as decimal logarithms  in Planck units. Horizontal scale extends approximately from the Planck time $t_P$ to almost the Hubble scale, $H_0^{-1}\approx 8\times10^{60} t_P$.
Field theory normally inhabits the entire right/lower half, but it is proposed here that in the lowest region, at large separations or small wavelengths, field modes are significantly entangled with geometry due to the Planck limit on directional information.   GeV  field modes are shown as an example: states are geometrically entangled beyond separations of a few kilometers.}
\end{figure}

The total amount of directional information---the number of  directional states  in a volume--- must not exceed the amount of directional information in the geometry.   
Directional entanglement of field states with geometry becomes significant when the smallest angular  structure of normal field states, on the  scale of the field wavelength $\Delta \theta\approx \lambda/z$, is smaller than the Planck angular resolution of the geometry, on scale $\Delta \theta_P\approx (l_P/ z)^{1/2}$.  That happens for systems of   duration  larger than a ``directional entanglement scale'' $\tau_g$,
\begin{equation}\label{validlimit}
 c\tau_g(\lambda)\approx \lambda^2/l_P.
\end{equation}
For larger systems--- in the lower  region shown in Figure (\ref{validity})--- field modes are significantly altered, due to a breakdown of the classical geometry approximation.
\subsection{Model of the Entangled System}

Suppose then that field modes and emergent geometry are entangled  parts of a composite quantum system. The entanglement is significant if there is not enough geometrical angular resolution to resolve the structure of the  field wave function, and transverse  coordinates do not behave classically.
   The paraxial decomposition  provides a way to quantitatively estimate the effect of separately quantizing the geometry in the transverse ($x,y$) directions, while retaining  the usual field-theory quantization of the longitudinal component.

Consider a wave function of the field mode (Eq. \ref{paraxialmodes})  entangled with  geometry, in a way similar to standard entanglement of quantum subsystems\cite{zeilinger1999,horodecki}. 
 The overall wave function can be approximated by a  product of field-like and geometry-like subsystems. 
The overall   states are modfied from those in classical geometry (Eq. \ref{paraxialmodes}) by entanglement with a directional wave function of the geometry:
\begin{equation}\label{product}
 \left[\sum_{k, {\cal K}, n}   \exp[i( k  z- \omega t-\phi)]  |A(k, n)\rangle |\psi_{{\cal K}f}(x-x_g,y-y_g;z)\rangle \right] \\
 \otimes  |\psi_{g}(x_g,y_g;z)\rangle 
 \end{equation}
where $f$ and $g$ refer to  the field and geometrical spatial wave functions, respectively.
Here, $x_g, y_g$ denote transverse geometrical position variables,
 and the wave function $|\psi_g(x_g,y_g,z)\rangle$ represents the amplitude for the geometry to depart from the  classical one by those values. This wave function for the geometry has a width approximately given by the minimal width for a Planckian wave in Eq. (\ref{rmin}),
\begin{equation}\label{planckwidth}
\sigma_g^2 (z)\approx  \langle \Delta r_{min}^2 (\lambda = l_P) \rangle =    l_P z/\sqrt{2}\pi.
\end{equation}

The product of the field and geometry structures gives an approximation to the overall transverse wave function of the combined system.
For a given $\lambda$, the directional uncertainty is field-like at  small $z$, geometry-like at large $z$.
The geometrical part $\psi_g$ represents a shared, coherent displacement for fields and massive bodies that only dominates at large separation $z$; in a small volume, there is no locally detectable deviation from usual theory. 

The nature  of the entanglement depends  on the scale,   orientation and preparation of the state.   The  phase of the overall state is modulated by the quantum effects of geometry  by roughly an amount
\begin{equation}
\Delta \phi\approx (\partial \psi_f/\partial x_i)\cdot \sigma_g,
\end{equation}
where $x_i$ with $i=1,2$ refers to the transverse coordinates  $x,y$. 
If this quantity is much less than unity, then entanglement has little effect, and the field behaves ordinarily, almost as if  geometry were classical; if it is much greater than unity, then the bulk  of the directional information (and uncertainty) is associated with the geometry.

For a mode with wavefronts nearly normal to the $z$ axis,  $|(\partial \psi_f/\partial x_i)|<< \lambda^{-1}$, 
so the  phase of modes close to this direction are  affected hardly at all by the geometry, even for $\tau>>\tau_{g}$. We can say that measurement of such a field state prepares or collapses the geometry into an eigenstate of this direction.  However, a   field mode oriented in a typical  direction has $|(\partial \psi_f/\partial x_i)|\approx \lambda^{-1}$, so entanglement typically changes the field phase  by
\begin{equation}\label{phase}
\Delta \phi\approx \sqrt{l_P\tau}/\lambda= \sqrt{\tau/\tau_g};
\end{equation}
so the field state is  substantially affected by geometry at $\tau>\tau_{g}$.
On the scale $\tau_{g}$, the quantum geometry typically changes the  phase of transverse modes by an amount of order unity, so the entanglement is substantial and the field modes no longer contain angular information approximately independent of the geometry or of each other.  
At small $\tau<\tau_{g}$, the geometry always produces  small fractional changes in the transverse field phases, so field theory behaves  almost as if it inhabits a classical geometry.  Small quantum-geometrical effects may nevertheless  be detectable in signals that measure small phase differences by using very large numbers of quanta, as discussed below for the case of laser interferometers.

The geometrical state has substantial spatial coherence.
Suppose a transverse measurement is made that ``collapses'' the geometrical position state associated with a particular direction to a definite value. The  geometrical wave function then spreads only slowly with time, with a width after time $\tau$ given by   $\sigma_g( z\approx c \tau)$.  Thus, neighboring bodies and particles share almost the same transverse geometrical state, if measured in the same direction, with positions differing by much less than their separation. In a continuous measurement in a fixed apparatus, the  difference nevertheless gives rise to a  slow fluctuation of measured transverse position of massive bodies.  The  slow time variation on a timescale $\tau\approx z/c$  has not been explicitly included here, as it represents a deviation from the paraxial approximation.

\subsection{Information content of geometry and field states}

In standard field theory, the the number of independent frequencies in a causal diamond of duration $\tau$ is roughly the  bandwidth divided by the resolution,  $\approx c\tau/\lambda$. The  number of  independent directional states comes from enumerating the complete family of independent  wave solutions in the volume of radius $c\tau$.  The  directional information grows like $(c\tau/\lambda)^2$, so the total information grows extensively, like the product ${\cal N}\approx (c\tau/\lambda)^3$.    The same result can be derived in the usual way using rectilinear coordinates and modes.

For  the quantum geometry, the radial information also grows linearly, like $\tau/t_P$. However,  in this case the    directional information  grows only like $\Delta\theta_P^{-2}\approx \tau/t_P$,  determined by the minimal angular uncertainty $\Delta\theta_P$ in the wave function of the fundamental Planck-wavelength gaussian mode. Therefore, the total geometrical information is the product of the radial and directional parts, about $(\tau/t_P)^2$. That  agrees with estimates from gravitational theory, for example with the entropy of black holes. 
 
As shown in Figure (\ref{degreesduration}), for large volumes and small $\lambda$ (but still $>> l_P$), the  geometrically-limited information is much less than predicted by standard field theory.  
 The information in fields in this regime   is  the product of the radial and limited directional information, 
${\cal N}\approx (c\tau/\lambda) (\tau/t_P).$ 
For smaller volumes  it is the other way around--- there are many  geometrical degrees of freedom not resolved by fields with  wavelengths much longer than Planck. As a result, the angular effect of geometrical entanglement cannot be resolved by fields in small scale experiments, so the geometry looks classical, and fields behave in the usual way.

\subsection{Connection with Path Integral Formulation}

Quantum states can also be described as path integrals.  The amplitude for a particle to be found at two places is a sum over all the amplitudes of the possible paths connecting them. A classical path is defined by Fermat's principle: the action integrated over a classical  path is an extremum, so to first order,  variations from the classical path lead to vanishing variations in phase. 

In terms of waves of a field, a classical path is thus everywhere normal to the wavefronts.  The position of a particle  is encoded in the phase of the field. The position in the along the path and in the transverse directions behave differently. Because the path is an extremum,  the phase in the transverse direction is quite insensitive to variations in position; it varies only in second order.  Thus, the phase change is small until the direction changes by the diffraction  limit for the wave frequency and propagation distance.

A model of spatial geometry based on a path integral over Planck frequency waves--- a ``space-time made of Planckian waves''---  gives the same directional uncertainty as the entangled paraxial model. The entanglement with field states gives a similar  directional uncertainty in the positions of particles much larger than a Planck length, of order $\sqrt{\tau/t_P}$.

This view gives some physical insight into the nature of spatial locality, coherence and entanglement.   Spatial position, as encoded in  phases of a Planck wavelength field, depends on the whole system that determines the local phase of the wave--- the past light cone of an event.   Events separate from  the path of a propagating particle   affect the Planck field phase, so the geometrical fluctuations of separate particle paths  vary coherently with no physical connection apart from their shared past.   

\section{Physical Effects in Systems Much Larger than the Planck Length}

Directional entanglement with geometry affects nonlocal aspects of field behavior on scales much larger than the Planck length.
Directional entanglement influences the behavior of physical systems both in near-vacuum states,  where fields propagate in nearly-flat space-time, and in situations with significant gravity. The model of directional entanglement presented above predicts an unusual combination of macroscopic effects,  which  are summarized here. 

\subsection{Particles} 

\subsubsection{Virtual Gravity of Field States}

 CKN\cite{cohen1999}  pointed out that the standard description of  field states is only valid up to a finite range, such that the sum of the energies of field states in a volume does not have more energy than a black hole of the same size, the densest configuration allowed by relativity. The energy density of a standard  field system with a mode frequency cutoff $m$ a and mean occupation $\bar n$ is about
 \begin{equation}\label{fielddensity}
\rho_f\approx (\bar n +1) m^4,
\end{equation}
independent of volume.   A black hole of size $L$  has a mean density $L^{-2}$ in Planck units, so 
for a system of size  $L$ and  field modes of mass $<m$, standard field theory with a UV cutoff at mass $m$ is in general only consistent with relativity in volumes smaller than 
\begin{equation}\label{fieldsystem}
L< L_{max}(m)= m^{-2}.
\end{equation}
  From a field theory point of view, this bound  must apply even to virtual states in flat space-time. 
To solve this problem, CKN posited an infrared cutoff on the extent of field states at this scale.  

Planckian directional entanglement as described here automatically implements  an IR cutoff at this scale (Eq. \ref{validlimit}), in a way that does not explicitly depend on $m$ or other detailed assumptions about field states. The proposal here effectively introduces the same IR limit on field states as CKN, but provides a specific geometrical rationale for  the  limit based on  geometrical directional information, and  predicts other kinds of specific macroscopic  effects on scales below the cutoff. 

 Taken literally, the vacuum energy density (Eq. \ref{fielddensity} with $\bar n= 0$) approaches the mass of a black hole as the size of a volume   $L\rightarrow L_{max}$. Another symmetry must be at work in a whole system to allow a global ground state that is nearly empty and nearly flat. This is discussed further below.

\subsubsection{Microscopic Interactions}
 
Consider interactions measured in direct particle experiments of the usual kind, such as particle  collider experiments. The interactions occur in a microscopic volume that is nevertheless much larger than the Planck length.
In this situation,
 geometry-limited paraxial modes are almost indistinguishable from standard plane waves; for example, at the TeV scale, the transverse phase gradient in the wave fronts is of the order of $10^{-8}/\lambda$.     The  precision of experiments, and the dynamic range of scales probed, are not sufficient to detect the geometrical constraint on directional information. 

\begin{figure}[t]
 \epsfysize=4in 
\epsfbox{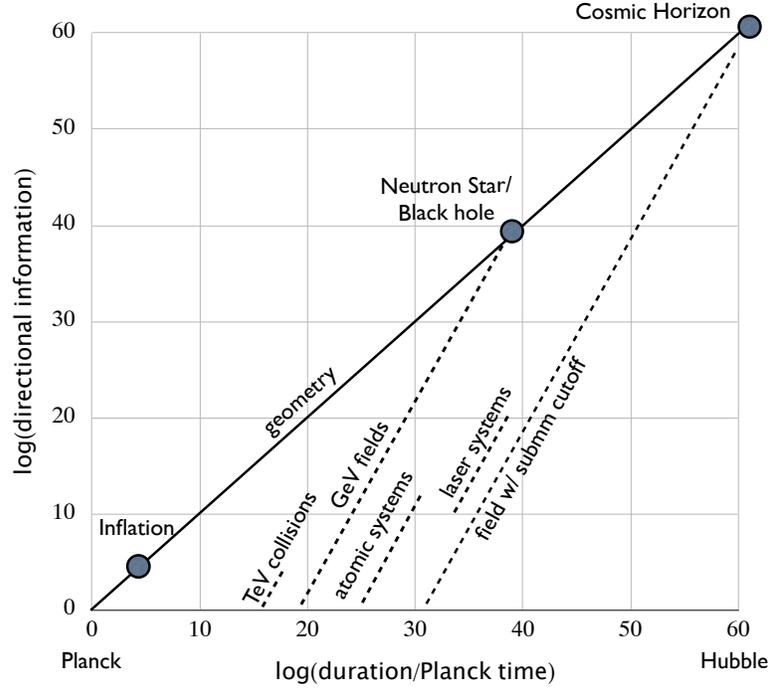} 
\caption{ \label{angularexamples}
Examples of directional information content of physical systems of various sizes. A saturated system of excited relativistic fields approximates the energy content and directional information of a black hole (albeit with less total information); these are represented by the neutron star/black hole dot. On the  cosmological scale, the    field vacuum, with a sub-millimeter cutoff enforced by directional entanglement, similarly matches  dark energy density and information.  Microscopic systems, such as hadronic collisions or assemblies of atoms,  are far from being limited by geometrical directional bounds, so field states act almost as if they are in a classical space-time.   Very small transverse field phase displacements caused by directional entanglement may nevertheless be detectable in large systems with very large numbers of coherent quanta, such as laser interferometers. }
\end{figure}

 CKN showed    that their IR cutoff is consistent with current tests using precision applications of field theory, based on  dimensional and renormalization group arguments.
Similar  estimates  apply to directional entanglement as well: in microscopic experiments, the  effects of directional entanglement are too small to appear in current particle experiments with realistic precision. 
 This point is made in Figure (\ref{angularexamples}) by the large vertical gap in directional information between  geometrical limits  and microscopic systems such as particle collisions and atoms.

\subsubsection{Tests of Lorentz Invariance Violation}

We have adopted a set of 
 coordinates and normal modes that introduce a preferred rest frame and a preferred axis. In this frame, Lorentz invariance is no longer manifest. However, the underlying relativistic wave equation describes the same Lorentz-invariant physical system as usual. Without the addition of the new Planck scale directional resolution limit, there is no physical difference and no violation of Lorentz invariance.

 On the other hand, Planckian directional entanglement  necessarily violates  Lorentz invariance.
 A measurement of geometry prepares a quantum state of the space-time with respect to a particular frame, fixed by the world-line that defines a system bounded by a causal diamond.
 However,  the violation only occurs in nonlocal measurements that are sensitive to transverse components of position or phase.  At the same time, the magnitude of the violation depends on system size, and is extremely small in systems much larger than the Planck scale.

 In the paraxial approximation used above,  the longitudinal part of the wave state is not changed  from the standard theory, that is,  the phase of a wave is not changed  along the propagation direction. 
For this reason, no effect (such as dispersion, or changes in time of flight) is predicted on particles that have traveled  over cosmic distances, as constrained by current experiments \cite{fermi2009,Laurent:2011he,HESS:2011aa}. This statement is independent of particle energy. 

The  amount of Lorentz invariance violation in the directions transverse to propagation is very small for systems much larger than the Planck length. Quantitatively, the amount of Lorentz invariance violation can be expressed as an effective transverse velocity. Over a distance $L$, the Planckian directional smearing is a transverse distance of about $L^{1/2}$ in Planck units, so the effective velocity is about $L^{-1/2}$.  On a typical laboratory scale, that is about $10^{-17}c$; on a cosmic scale, it is about $10^{-30}c$. These effects lie well below the threshold of experimental limits.

\subsection{Interferometers}

The most sensitive technique for measuring very small phase differences in macroscopic systems is Michelson interferometry. In an interferometer, modes of  light propagating in different directions can be mixed by reflecting surfaces of macroscopic massive bodies--- in particular,  beamsplitter mirrors. The signal depends on the phase difference of light impinging on the beamsplitter from two directions, so it is sensitive to  variations in transverse position.  The standard quantum theory of interferometers\cite{caves1980,caves1980b}  includes  entanglement of  mirrors and laser fields as subsystems in a classical geometry, and predicts a standard quantum noise limit.  Planckian directional entanglement adds a new source of noise, that depends only on the spatial configuration of an interferometer.   

For an apparatus of size $L_a$, the quantum geometry produces fluctuations on timescale $\tau\approx L_a$, corresponding to fluctuations   in phase with amplitude  on the order of Eq. \ref{phase}.   The  fluctuations in displacement  and angle,   $\Delta x_\perp \approx  \Delta \theta L_a\approx \Delta \phi \lambda$ have a spectral density, or variance  per  frequency interval of the fluctuations,  given approximately by the Planck time:
\begin{equation}\label{planckdensity}
 d \langle \Delta\theta^2 \rangle / df  \approx t_P = 5.39 \times 10^{-44} {\rm Hz}^{-1},
\end{equation}
with most of the fluctuation in an apparatus of size $L_a$ coming at frequencies $ f \approx c/L_a$.
 These predicted phase fluctuations may be detectable, and distinguishable from other sources of noise by their correlations in space and time\cite{Hogan:2012ne,Hogan:2010zs,Hogan:2012ib}. 

\subsection{Vacuum Energy Density}

The most spectacular experimental failure of standard field theory appears in cosmology: it vastly over-predicts the gravitational energy density of field vacuum states in the   cosmic system,  compared with the effective  mean density of the ``dark energy'' associated with the acceleration of cosmic expansion \cite{weinberg89,Frieman:2008sn}, even though it correctly predicts behavior of vacuum fluctuations  in laboratory systems\cite{lamoreaux2005,Jaffe:2005vp}.  This contradiction is an extreme example of that discussed above between the density of virtual field states and black holes, for   a system size fixed by the cosmic expansion rate $H$. The  mean cosmic density is about $H_0^2\approx 10^{-122}$ in Planck units, and is much smaller than Eq. (\ref{fielddensity})  for scales $m$ of the Standard Model.
 The proposal here does not explain the value of dark energy, but it does resolve
this extreme contradiction.

   The  number of  field degrees of freedom ${\cal N}$ is the product of radial and directional information. For field modes with  frequency  $<m$  directionally entangled with geometry in a volume of size  $\tau$,   ${\cal N}\approx  m \tau^2$. The  vacuum energy  density of states with mean occupation $\bar n=0$  is then
$\rho_{vac}\approx {\cal N}m/\tau^3 \approx   
m^2 / \tau.$
We should  only count the density of states for which directional entanglement is small from Eq. (\ref{validlimit}),
$\tau<  m^{-2}$.    Any  choice of $\tau$ and $m$ then gives $
\rho_{vac}<   \tau^{-2},$ so the field density is at most comparable with  the mean density of matter in a  system of  gravitational curvature radius $\tau$.  Thus,  
 directional entanglement  solves the main  part of the vacuum  energy problem: it eliminates the extreme fine tuning.

This estimate, if taken literally, also predicts that fields even in the   vacuum state  may approximately saturate this bound for a volume of any size, which is clearly not the case.  Some other symmetry or equilibration principle must be at work in the whole quantum system of fields and geometry, to allow large volumes of space that are not strongly gravitationally curved, but almost flat. Furthermore, the (tiny) value of the cosmological constant is determined by some  mechanism that breaks that principle. The entanglement hypothesis does not determine  that value, although it may  connect it to some microscopic scale of the field theory.
\subsection{Black Holes }

\begin{figure}[t]
 \epsfysize=4in 
\epsfbox{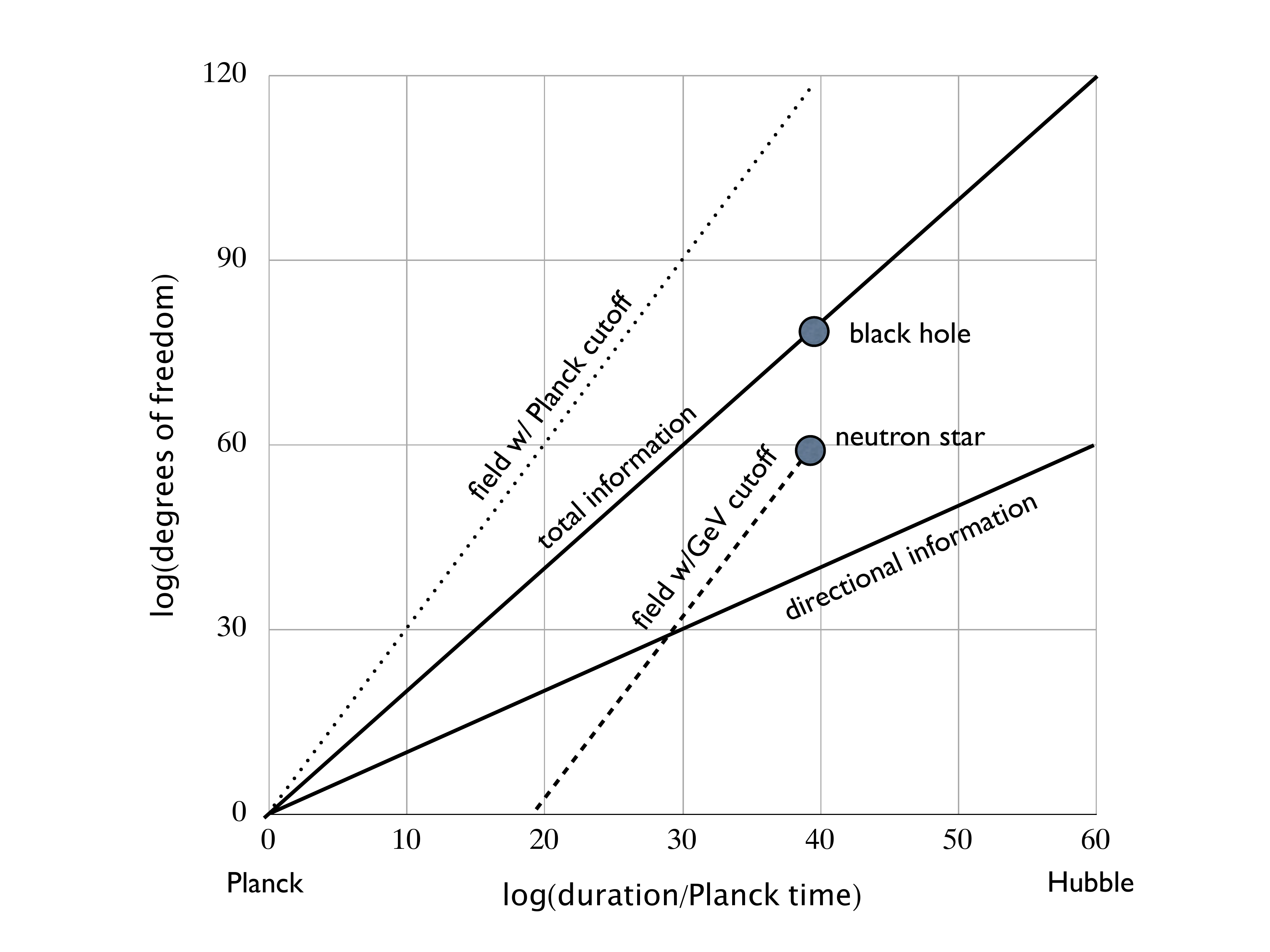} 
\caption{ \label{degreesduration}
Information, or number of degrees of freedom, as a function of the duration  of a space-time causal diamond in Planck units.  Solid lines show total  information and  directional information. 
Dotted line shows  standard field theory in a classical geometry, for a UV cutoff at the  Planck scale. 
Dashed line shows standard  fields with a cutoff at the GeV scale, but terminated on the scale of directional entanglement with geometry.
A  large system is needed before the geometrical entanglement significantly constrains standard field degrees of freedom. 
Large dots indicate the geometrical  information in a stellar mass black hole, and the (far smaller) GeV-scale field information in a neutron star. The gap between them corresponds to the large increase in entropy that occurs when an event horizon forms.}
\end{figure}

The approximations used here break down when the curvature of wave fronts is comparable to the curvature of emergent space-time, so they cannot be applied to black holes.
However, the  maximum total information and  energy in a region  roughly match to  black hole states.  For any $\lambda$,  a thermally populated field state in a volume size $\approx \tau_g(\lambda)$, or degenerate relativistic fermonic matter with occupation number of order unity, has approximately the mass and radius of a black hole of the same size. For example, a nearly relativistic, self-gravitating degenerate system with Fermi energy on the GeV scale corresponds roughly to a neutron star.  It  has a size of about $\tau_g(\lambda\approx 1/{\rm GeV}) $, of order a few kilometers.
The black hole configuration of  comparable mass and size corresponds to  far  higher entropy. Presumably, in that configuration the geometrical degrees of freedom are  in a different  state. In the black hole state, the classical directional information is completely scrambled by strong gravity near the hole.   The gap (in figure \ref{degreesduration}) between neutron star and black hole represents   the much larger number of  geometrical states compared to field states.   If the formation of  a black hole   resembles a phase transition\cite{Dvali:2012en,martinec}, the effective strong-field geometrical degrees of freedom could have  a very different character from those of nearly-flat space.

\acknowledgments

I am grateful  for the hospitality of the Aspen Center for Physics, which is supported by  National Science Foundation Grant No. PHY-1066293.  This work was supported by the Department of Energy at Fermilab under Contract No. DE-AC02-07CH11359.

\vfil
\break

 \end{document}